# Ultrafast dynamics of atomic correlated disordering in photoinduced $VO_2$

Wen-Hao Liu[1]†, Feng-Wu Guo[1,2]†, Lin-Wang Wang[1]*, and Jun-Wei Luo[1,2]*

[1]*State Key Laboratory of Superlattices and Microstructures, Institute of Semiconductors, Chinese Academy of Sciences, Beijing 100083, China*

[2]*Center of Materials Science and Optoelectronics Engineering, University of Chinese Academy of Sciences, Beijing 100049, China*

†These authors contributed equally to this work.

*Email: lwwang@semi.ac.cn; jwluo@semi.ac.cn

**Abstract:**

Recent experiments suggest that atomic disordering dynamics are more universal than conventional coherent processes in photoinduced phase transitions (PIPTs), yet its mechanism remains unclear. Using real-time time-dependent density functional theory (rt-TDDFT), we find that, at lower photoexcitation, higher lattice temperature accelerates atomic disordering, which thereby lowers the threshold for phase transition, by thermally exciting more phonons to randomize the lattice vibrations in $VO_2$. Above this threshold, however, we observe that the transition timescale and atomic disordering become temperature-independent since thermally excited lattice vibrations induce a similar evolution of photoexcited holes. Additionally, we show that photoexcitation initially elongates the V-V dimers followed by a rotation with tangential displacements (along the $z$-axis) mediated by O atoms, resulting in strongly correlated motion along the $z$-axis. Consequently, atomic disordering is more dominant along the $x$ direction, attributed to the relatively unrestricted motion of V-V dimers along this direction. The motion of V atoms along the $z$-axis is more constrained, leading to less disorder along the $z$-axis, which results in a “correlated disorder” phenomenon. This anisotropic disordering in $VO_2$ offers new insights into PIPTs mechanisms, guiding future studies on photoinduced disordered transitions.

The conventional understanding of ultrafast photoinduced phase transitions (PIPTs) typically characterizes them as atom collective motion along a potential energy surface (PES) or as coherent atomic oscillations resulting from photoexcited coherent phonons [1-10]. Therefore, prevailing theoretical models often classify ultrafast dynamics as coherent processes [11-14]. However, a diffuse scattering experiment [15], unlike typical X-ray or electron diffraction approaches that measure averages across numerous unit cells [16-19], revealed atomic disordering during PIPTs rather than a collective coherent motion in $VO_2$. Furthermore, recent research suggests that the disordering and timescale of photoinduced phase transition in $VO_2$ are independent of the initial lattice temperature, attributed to the presence of non-conservative forces resulting exclusively from electron-phonon collisions [10]. Our previous studies have elucidated that the competition between thermal-induced and photoinduced atomic driving forces serves as a fundamental mechanism for the dynamics of atomic disordering and coherence in the photoinduced phase transition of $VO_2$ [20].

Photoinduced disordering dynamics are commonly considered as resulting from spatial uncorrelation [21]. Interestingly, a recent experiment [22] has observed that atomic displacements maintain a degree of correlation during the disordered phase transition in $VO_2$ — a phenomenon termed "correlated disorder" [23], where atomic motion remains correlated along one dimension but disordered along others. However, the underlying mechanisms of correlated versus uncorrected disorder during PIPTs are not yet well understood. Moreover, the influence of external factors, such as temperature and laser fluence, on atomic disordering dynamics remains unclear.

In this letter, we employ real-time time-dependent density functional theory (rt-TDDFT) to investigate the influence of initial lattice temperature and laser fluence on PIPTs in $VO_2$. Our findings show that, at lower laser fluence, initial lattice temperature plays a significant role in PIPTs, with higher temperatures accelerating atomic disordering and reducing the phase transition threshold. However, as laser fluence increases, atomic disordering and phase transition timescale become independent of the initial lattice temperature. This temperature independence arises from the lattice vibration-induced charge transfer, which results in distributions of photoexcited holes being similar at different temperatures. Furthermore, we demonstrate that upon photoexcitation, the V-V dimers initially elongate along the $x$ direction and then rotate along the $z$ direction, with the O atoms acting as a bridge. As a result, atomic motion along the $z$-axis exhibits correlated behavior, leading to smaller disordering along the $z$-axis than the $x$-axis. The revealed atomic anisotropic disordering explains the "correlated disorder" observed in $VO_2$ and provides critical insights into the mechanisms of the PIPTs.

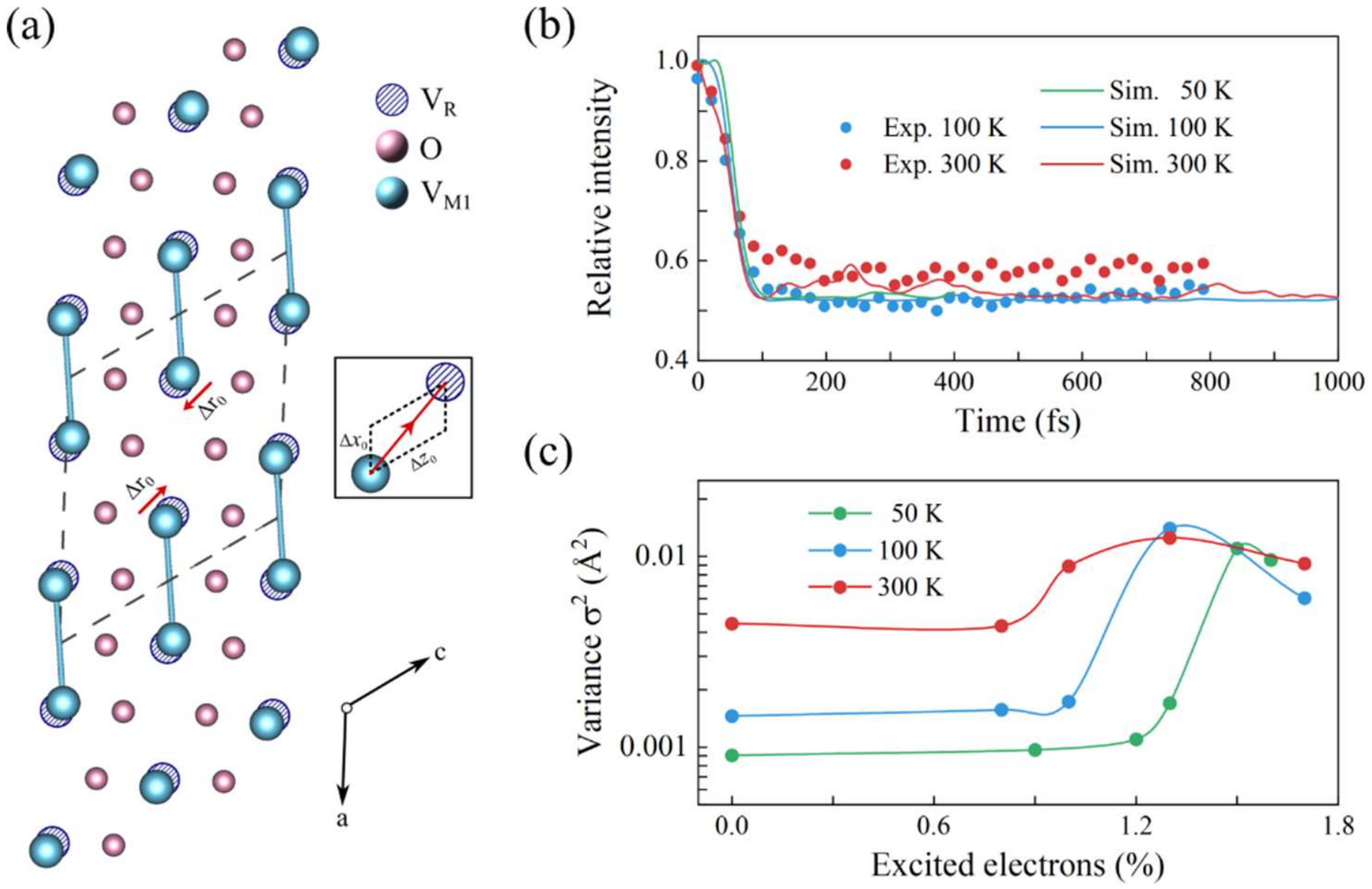

**FIG. 1. Photoinduced atomic disordering.** (a) Schematic illustration of the M1 phase structure with solid blue spheres representing V atoms and the R phase structure with blue striped spheres representing V atoms. All O atoms are denoted pink spheres. The dashed parallelepipeds indicate the unit cells of the M1 phase. The inset shows an enlarged view of the V atom positions in the M1 phase and R phase. (b) Temporal evolution of simulated Bragg peak intensity for the ($\bar{1}22$) reflection under 1.6% electronic excitations for an initial temperature of 50 K (green line) and 1.3% electronic excitations for 100 K (blue line) and 300 K (red line). The blue and red dots represent the relative intensity of ($\bar{1}22$) peak induced by a laser fluence of 46.5 mJ/cm$^2$ at 100 K and 28.0 mJ/cm$^2$ at 300 K, respectively, extracted from experimental data [10]. (c) Variance of atomic displacement as a function of the electronic excitation percentage ($\eta$) at 50 K, 100 K, and 300 K.

To investigate the photoinduced correlated disorder observed in $VO_2$, we performed rt-TDDFT simulations utilizing the PWmat package [24], employing the local density approximation functional (LDA) + U exchange-correlation with Hubbard U = 3.4 eV. The LDA+U method has been widely applied in $VO_2$ [15,19,25-27]. Our simulation model consists of a 96-atom supercell, with a plane-wave basis chosen with a cutoff energy of 45 Ry and a 2×2×2 Monkhorst-Pack k-point mesh for Brillouin-zone sampling. To obtain the thermodynamically equilibrated lattice structures under different temperatures (50 K, 100 K, and 300 K), we conducted BOMD simulations using a time step of 1 fs [Supplementary Materials Fig. S1 [28]]. After that, we consider these temperature-dependent lattice structures as the initial structures for our rt-TDDFT simulations under different initial lattice temperatures. In the rt-TDDFT dynamic simulations, we employed a time step of 0.1 fs to guarantee convergence for large simulation timescale. To mimic the laser pulse photoexcitation, we apply a time-dependent external electric field in a Gaussian shape,

$$E(t) = E_0 \cos(\omega t) \exp[-(t-t_0)^2/(2\sigma^2)], \quad (1)$$

where we set photon energy $\hbar\omega$ = 1.55 eV, time delay of $t_0$ = 25 fs, and width parameter $\sqrt{2}\sigma$ = 10 fs. The electric field $E_0$ is adjusted to reproduce the electronic excitations by photoexcitations [Supplementary Materials Fig. S2 [28]].

Figure 1(a) shows that upon photoexcitation, $VO_2$ undergoes a transition from the insulating M1 phase (*P*21/*C*), characterized by twisted V-V dimerization and stablized at low temperature, to the metallic R phase (*P*42/*mnm*), with V atoms exhibiting displacements of $\Delta x$ = 0.083 Å and $\Delta z$ = 0.145 Å [19], and the phase is stablized above ~ 340 K [29]. Specifically, our rt-TDDFT simulations can achieve the photoinduced M1-to-R phase transition in $VO_2$ at different temperatures and laser fluences [Figs. 1(b), 1(c) and Supplementary Materials Figs. S3, S4 [28]].

To examine the atomic fluctuations and disordering as observed in diffuse scattering experiments [10,15], we employ the average variance ($\sigma^2$) of atomic displacements, defined as:

$$\sigma^2(t) = \langle (u(t) - \langle u(t) \rangle)^2 \rangle, \quad (2)$$

where $u(t)$ represents the time-dependent V-V bond lengths [Supplementary Materials Fig. S3]. Figure 1(c) shows the average variation $\sigma^2$ within 400 fs following the photoexcitation as a function of laser fluence (photoexcitation strength) η for intial lattice temperatures 50 K, 100 K, and 300 K [Supplementary Materials Fig. S5 [28]]. Under lower photoexcitation, the variance $\sigma^2$ sustains relatively stable, with the atomic disordering being $\sigma^2$(300 K) > $\sigma^2$(100 K) > $\sigma^2$(50 K) [Fig. 1(c)]. As photoexcitation approaches the phase transition threshold, atomic disordering increases sharply, consistent with our previous report [20]. The atomic disordering $\sigma^2$ varies most significantly at 1.0% photoexcitation for an initial lattice temperature 300 K, 1.3% photoexcitation for 100 K, and 1.5% photoexcitation for 50 K, which corresponds to the threshold of photoinduced phase transition [Supplementary Materials Fig. S3 [28]]. Our results also confirm that the laser threshold is influenced by the initial lattice temperature, with lower temperatures requiring higher laser fluence to achieve the phase transition, which is consistent with the experimental results [30].

Interestingly, when laser fluences above these phase transition thresholds, the laser-induced atomic disordering $\sigma^2$ becomes comparable at different temperatures. At these investigated temperatures [Fig. 1(b) and Supplementary Note 2 [28]], the time-dependent relative diffraction intensity $I(t)/I(t=0)$ from the ($\bar{1}22$) reflection drops at almost a same time following the photoexcitation to a minimum, indicating that the timescale of the photoinduced M1-to-R phase transition is independent of the initial lattice temperature, in excellent agreement with a recent experimental measurement [10]. To unravel the underlying mechanism, we analyze the number and distributions of photoexcited holes spatially on different V-V dimers at 20, 50, and 200 fs following photoexcitation [Figs. 2b and 2d, Supplementary Materials Fig. S6 [28]]. Initially, the incident photons excite valence electrons from the bonding states in the valence band to the antibonding states in the conduction band, leaving holes in the valence band. The modulation of carriers between bonding and antibonding states generates interatomic forces to elongate the V-V dimers to lower the total energy. Figure 2d shows that the photoexcited holes are nearly uniformly distributed across the V-V dimers at 20 fs, and, thus, the generated interatomic forces on all dimers are equivalent in magnitude. However, the thermally excited lattice vibrations could induce random movements of the V atoms, leading to fluctuations in the V-V bond lengths across V-V dimers [Fig. 2a] and, in turn, causing band energy variations spatially. The photoexcited electrons (holes) will move toward the local band energy minimum (maximum), which thus acts as a trapping center for the excited electrons and holes, as shown in Fig. 2c, and breaks the equivalence in the photogenerated interatomic forces. The localization of photocarriers further amplifies the transient atomic disordering induced by the phonon-induced lattice vibrations. Figure 2a indeed shows that, as the system progresses to 50 fs, the atomic disordering begins to increase. This increase is also reflected in the spatial distribution of photoexcited holes, which becomes more inhomogeneous [Figs. 2b and 2d]. Furthermore, we find that the dynamic processes of atomic disordering at all investigated initial lattice temperatures are similar, as shown in Supplementary Materials Figs. S7 and S8 [28], indicating that the final degree of atomic disordering becomes independent of the initial temperature.

To validate the phonon-assisted self-amplification process, we consider an inhomogeneous initial state with a pre-existing polaron in $VO_2$ [Supplementary Materials Fig. S9 [28]]. In this case, local distortion of the pre-existing polaron, instead of thermal-induced lattice vibrations, provides the initial seed for the self-amplified localization of photoexcited holes across the V-V dimers, breaking the equal magnitude of photogenerated interatomic forces acting on the V-V dimers, which aligns with our proposed mechanism [Fig. 2c]. As a result, the V atoms associated with the pre-existing polaron remain in a disordered state throughout the entire process [Supplementary Materials Fig. S10 [28]]. Notably, the final degree of atomic disordering in the initially inhomogeneous system closely resembles that of the initially homogeneous state. This suggests that, although the different sorts of local distortions may influence the initial charge distribution, they do not significantly alter the overall dynamics of atomic disordering during the PIPTs.

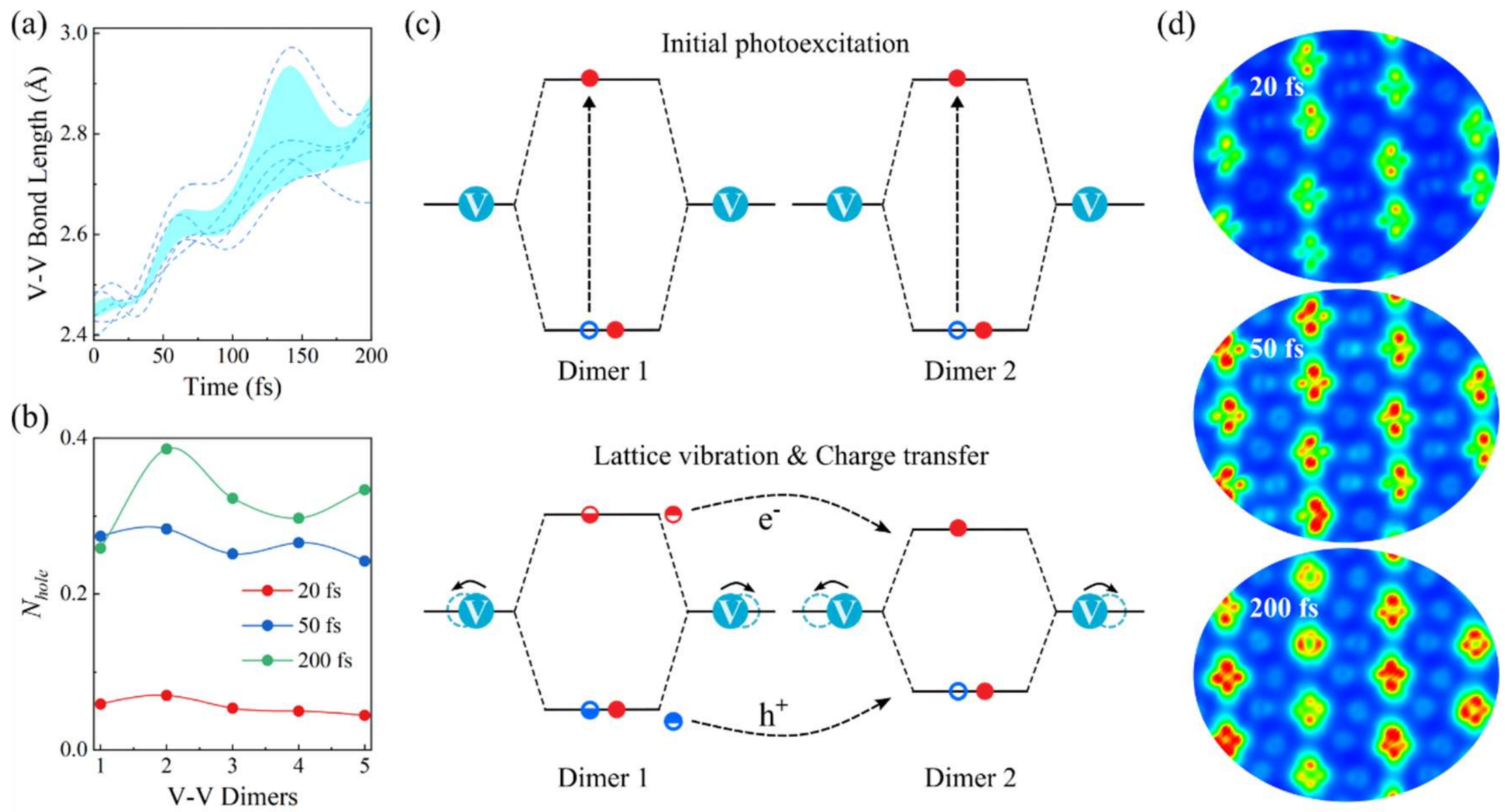

**FIG. 2. Mechanism of photoinduced atomic disordering.** (a) Temporal evolution of V-V dimerized bond lengths for 1.3% photoexcitation at 100 K with light blue shaded regions representing the fluctuation of the V-V bond lengths. (b) Photoexcited holes on V-V dimers within -1.6-0 eV below the Fermi Level at 20 fs (red), 50 fs (blue), and 200 fs (green). (c) A schematic diagram illustrating the process of initial photoexcitation, followed by thermal vibration-induced charge transfer. Red circles represent photoexcited electrons, blue circles indicate photoexcited holes, and light blue circles depict the movement of V atoms. (d) Distributions of the photoexcited holes on the $(0\bar{1}1)$ plane at 20 fs, 50 fs and 200 fs following photoexcitation.

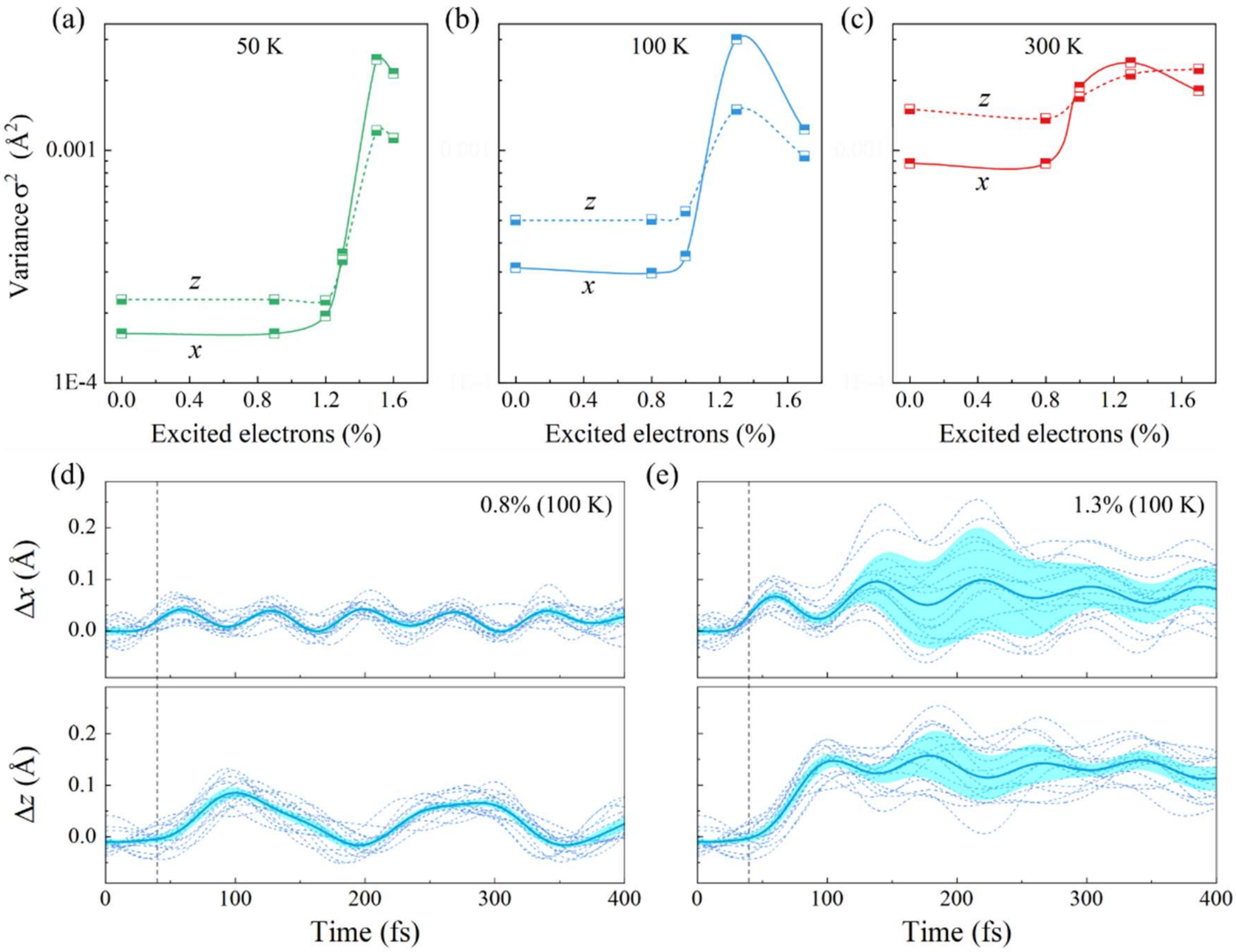

**FIG. 3. Directional dependence of atomic disordering.** The variance of atomic displacement along the *x* and *z* axes as a function of the percentage (η) of electronic excitation at (a) 50 K, (b) 100 K, and (c) 300 K. Temporal evolution of atomic displacement along the *x* and *z* axes for (d) 0.8% photoexcitation and (e) 1.3% photoexcitation at 100 K. The light blue shaded regions in (d, e) indicate the fluctuation of atomic disordering. The grey dashed lines in (d, e) represent the starting time of atomic displacement along the *z*-axis.

Futhermore, we perform calculations of atomic disordering degree along the *x*- and *z*-axes using Equation (2) to assess the directional dependence of atomic disordering [Fig. 3 and Supplementary Materials Fig. S11 [28]]. Figures 3(a-c) illustrate the average variation along the *x* and *z* directions within 400 fs as a function of photoexcitation strength η at initial lattice temperatures of 50 K, 100 K, and 300 K. At lower photoexcitation, the variance $\sigma^2$ along the *z*-axis is larger than that along the *x*-axis [Figs. 3(a-c)], attributed to the larger displacement of V atoms along the *z*-axis compared to the *x*-axis [Fig. 3(d) and Supplementary Materials Fig. S12 [28]]. Surprisingly, as photoexcitation increases to the phase transition threshold, atomic disordering along the *x*-axis rapidly exceeds that along the *z*-axis [Figs. 3(a-c) and 3(e)]. This indicates that atomic disordering becomes more prominent along the $x$-direction, despite the $z$-axis displacement of V atoms remaining larger than that along the $x$-axis [Fig. 3(e) and Supplementary Materials Fig. S12 [28]]. These findings emphasize the strong directional dependence (anisotropy) of atomic disordering during the phase transition.

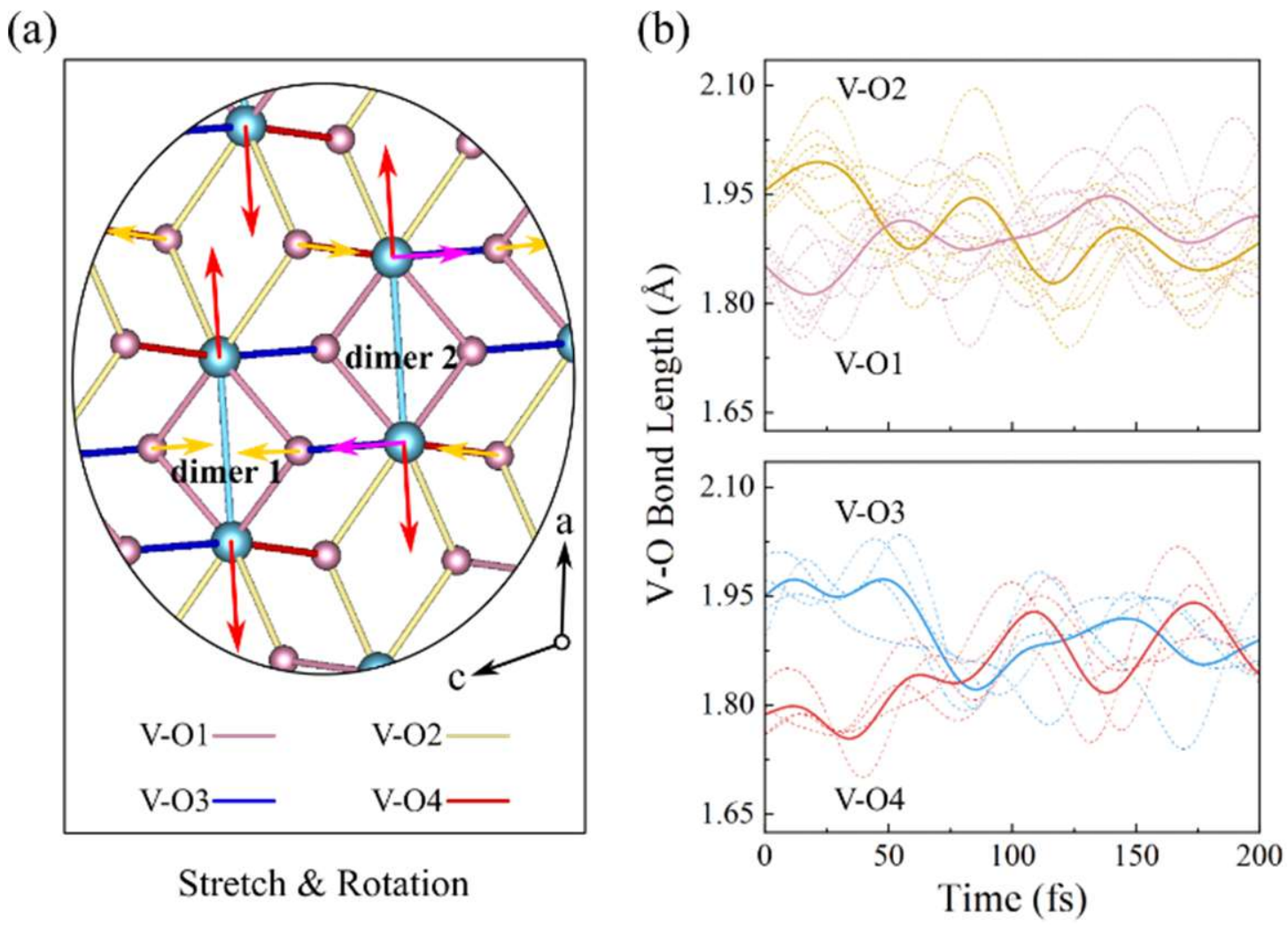

**FIG. 4. Atomic correlated disordering.** (a) Schematic diagram of the photoexcited force during stretching and rotation. Red arrows indicate the stretching forces acting along the V-V dimers ($a$-axis). Yellow arrows represent forces on O atoms arising from the elongation of V-V short bonds and the contraction of long bonds, which collectively induce tangential atomic forces (pink arrows) on the V-V dimers. (b) Temporal dynamics of various V-O bond lengths. The dashed lines represent individual V-O bond lengths, and the solid lines indicate the average V-O bond lengths. The V-O bonds in $VO_2$ are categorized into four types: V-O1 (pink), V-O2 (orange), V-O3 (blue), and V-O4 (red), as shown in both (a) and (b).

To understand the atomic anisotropic disordering in $VO_2$, we first analyze atomic displacements along the $x$- and $z$-axes during the PIPTs. The displacements along the $x$-axis are attributed to the elongation of V-V dimers, while those along the $z$-axis are caused by the rotation of the V-V dimers, as schematically shown in Fig. 4b. We find that the displacement of V atoms along the $x$-axis begins to increase around 45 fs, followed by an increase of the displacement along the $z$-axis at approximately 80 fs [Figs. 3(d) and 3(e)], indicating that the V-V dimers undergo the elongation initially followed by rotation [Fig. 1(a)]. This finding is in excellent agreement with the previous experimental observations [17].

We illustrate that the elongation-rotation process in the photoinduced dissociation of the V-V dimers is closely tied to the V-O bonds, a relationship often overlooked in literatures [16-19].

Along with the stretching of the V-V dimers (and the shortening of the longer V-V bond) along the *x*-axis driven by photogenerated atomic forces, the two diaognal O atoms in a rhombus formed with each V-V dimer (called dimer-1) receive opposite forces along the *z*-axis to pull them closer due to the positive Poisson ratio [Fig. 4(b)]. The horiztonal force to pull the right side O atom will also pull its bonded V atom, which is belong to another V-V dimer (called dimer-2), to move along the *z*-axis (i.e., from right to left). Meanwhile, the shortening of the longer V-V bond leads to two diagonal O atoms of a rhombus formed with it recive opposite forces to push them far away along the *z*-axis due the positive Poisson ratio. The horiztonal force acting on the right-side O atom also pushes the V atom of dimer-2 to move from left to right along the *z*-axis, giving rise to clockwise rotation of dimer-2, in combination with the movement of its other V atom from right to left. This analysis can be extended to all V-V dimers, where the motion of V atoms along *z*-axis is constrained by the displacement of V atoms along the *x*-axis and the surrounding O atoms.

These atomic forces are also reflected in the evolution of the V-O bonds [Fig. 4(b)]. As a result, the stretching of the V-V dimers and the shortening of the longer V-V bonds elongate the V-O1 bonds and shorten the V-O2 bonds, causing both to converge to an equal length of ~ 1.9 Å at 45 fs [Fig. 4(a)]. Subsequently, the forces exerted on the V-O3 and V-O4 bonds drive their equalization to ~ 1.9 Å at 80 fs [Fig. 3(a)], initiating the rotational motion of the V atoms along the z-axis. Notably, distinguishing between elongation and rotational processes in $VO_2$ via phonon measurements is challenging, as both produce similar phonon signatures [Supplementary Materials Fig. S13]. This contrasts with other systems, such as indium nanowires on silicon surfaces, where these processes exhibit distinct phonon characteristics [31,32].

To further strengthen the effect of atomic correlated disordering, we froze the positions of the V atoms along the *x* and *z* directions, respectively. These results reveal that freezing the *x*-axis position restricts V atomic movement along the *z* direction, while freezing the *z*-axis position does not limit V atomic movement along the *x* direction [Supplementary Materials Fig. S14 [28]]. This indicates that atomic motion along the *z*-axis is more correlated, resulting in lower disorder along this direction than along the *x*-axis. Our findings also provide a mechanistic explanation for recent experimental observations in $VO_2$ [22], which reported the formation of photoinduced local polarons along the *z*-axis. This correlated effect along the *z*-axis facilitates the emergence of these polarons, further supporting the directional dependence of disordering in the PIPTs.

In summary, our research reveals that the initial lattice temperature has a pronounced effect on atomic disordering and the phase transition threshold under lower photoexcitation. However, once the laser fluence surpasses phase transition threshold, both the degree of atomic disordering and the timescale of the phase transition become independent of the initial temperature. This temperature independence results from similar non-uniform distributions of photoexcited holes at various temperatures. Additionally, we reveal an anisotropic atomic disordering of V atoms in photoinduced $VO_2$. Upon reaching the phase transition threshold, disorder along the $x$-axis becomes more pronounced than along the $z$-axis. This anisotropy arises from correlated atomic motion along the $z$-axis, where the initial elongation of V-V dimers is unrestricted along the $x$-axis but is followed by a rotation constrained by the interaction with O atoms along the $z$-axis. These insights shed light on the mechanisms driving atomic anisotropic disordering during the PIPTs in $VO_2$. Our findings provide valuable insights for future studies exploring the photoinduced disordered transition path in other materials.

## ACKNOWLEDGMENTS

The work was supported by the National Natural Science Foundation of China (NSFC) under Grant Nos. 11925407, 61927901, 12474076 and 12204470, CAS Project for Young Scientists in Basic Research under Grant No. YSBR-026.

Supplemental Materials for

# Ultrafast dynamics of atomic correlated disordering in photoinduced $VO_2$

Wen-Hao Liu[1]†, Feng-Wu Guo[1,2]†, Lin-Wang Wang[1]*, and Jun-Wei Luo[1,2]*

†These authors contributed equally to this work.

*Email: lwwang@semi.ac.cn; jwluo@semi.ac.cn

**Note 1: BOMD simulations and RT-TDDFT simulations**

We conducted BOMD simulations over a 1 ps timespan within a time step of 1 fs to obtain structures at 50K, 100 K, and 300 K. The temporal evolution of temperatures for various examples is depicted in Fig. S1. Notably, The temperature exhibits minor periodic fluctuations over time, indicating that the lattice remains stable at this temperature, signifying thermodynamic equilibrium.

Furthermore, we consider these thermodynamically equilibrated lattice structures at different temperatures as the initial structures for our subsequent rt-TDDFT simulations. In the rt-TDDFT dynamic simulations, we employed a time step of 0.1 fs. To simulate photoexcitation, we introduced the A-field into the *k*-space of the Hamiltonian.

$$H(t) = 1/2\big(-i\boldsymbol{\nabla} + \boldsymbol{A}(t)\big)^2 \tag{1}$$

where,

$$\boldsymbol{E}(t) = -\frac{\partial \boldsymbol{A}(t)}{\partial t} \tag{2}$$

In our rt-TDDFT simulations, we apply a time-dependent external electric field in a Gaussian shape,

$$E(t) = E_0\, cos(\,\omega t)\, exp[\,-(t - t_0)^2/(2\sigma^2)], \tag{3}$$

where $E_0$ is the amplitude of the electric field, photon energy $\hbar\omega$ = 1.55 eV, a time delay of $t_0$ = 25 fs and a width parameter $\sqrt{2}\sigma$ = 10 fs. The electric field $E_0$ is adjusted from 0.35 ~ 1.60 × $10^{12}$ W/cm$^2$ to attain different electronic excitations at different tempeartures [Fig. S2].

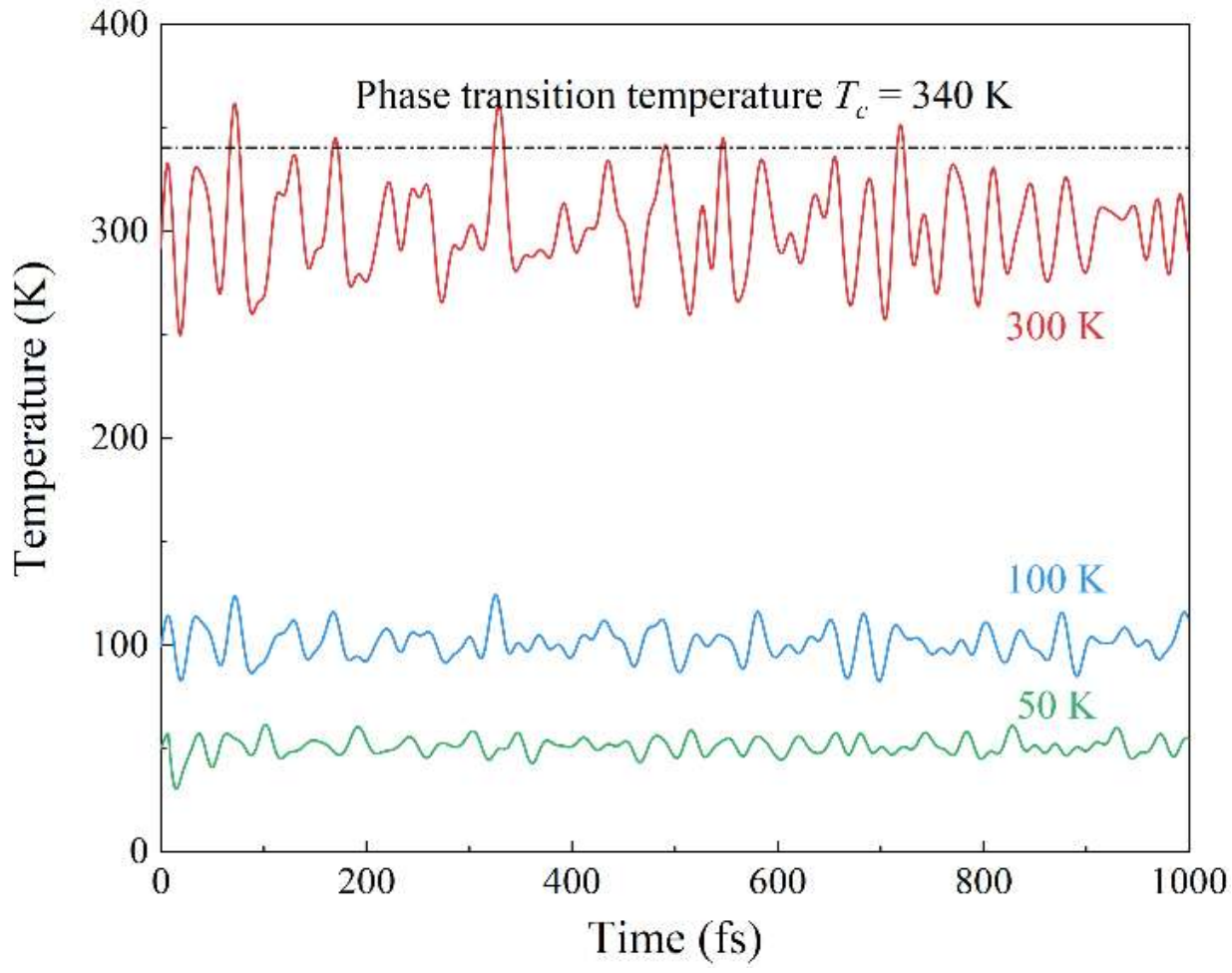

FIG. S1. Lattice temperature as a function of time in BOMD simulations. The dashed line represents the phase transition temperature from M1-to-R phase.

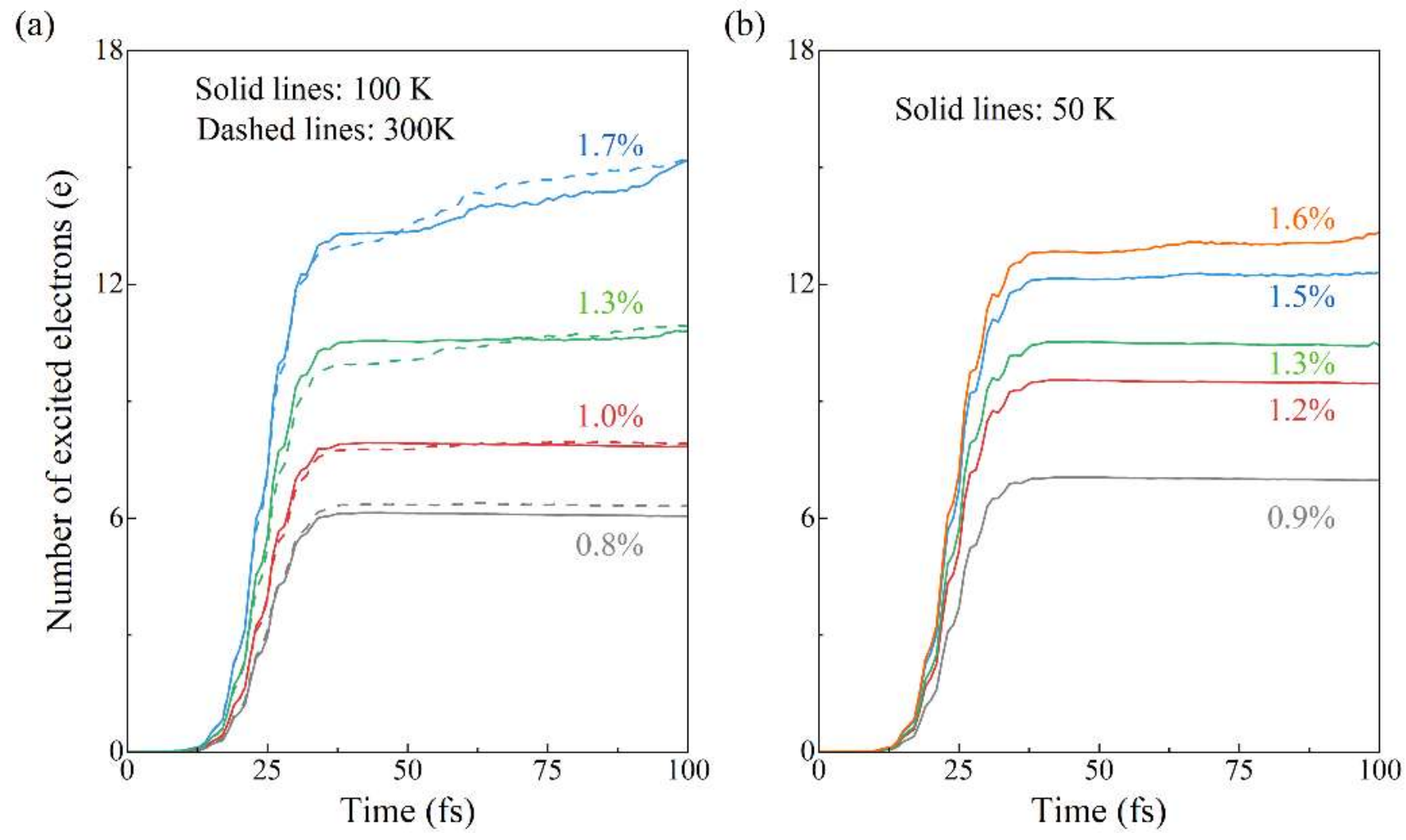

FIG. S2. Evolution of the number of excited electrons from the valence bands to conduction bands upon different photoexcitation at 50 K, 100 K, and 300 K. The proportion of excited electrons to the total number of electrons is defined as η.

## Note 2. Photoinduced phase transition and atomic disordering

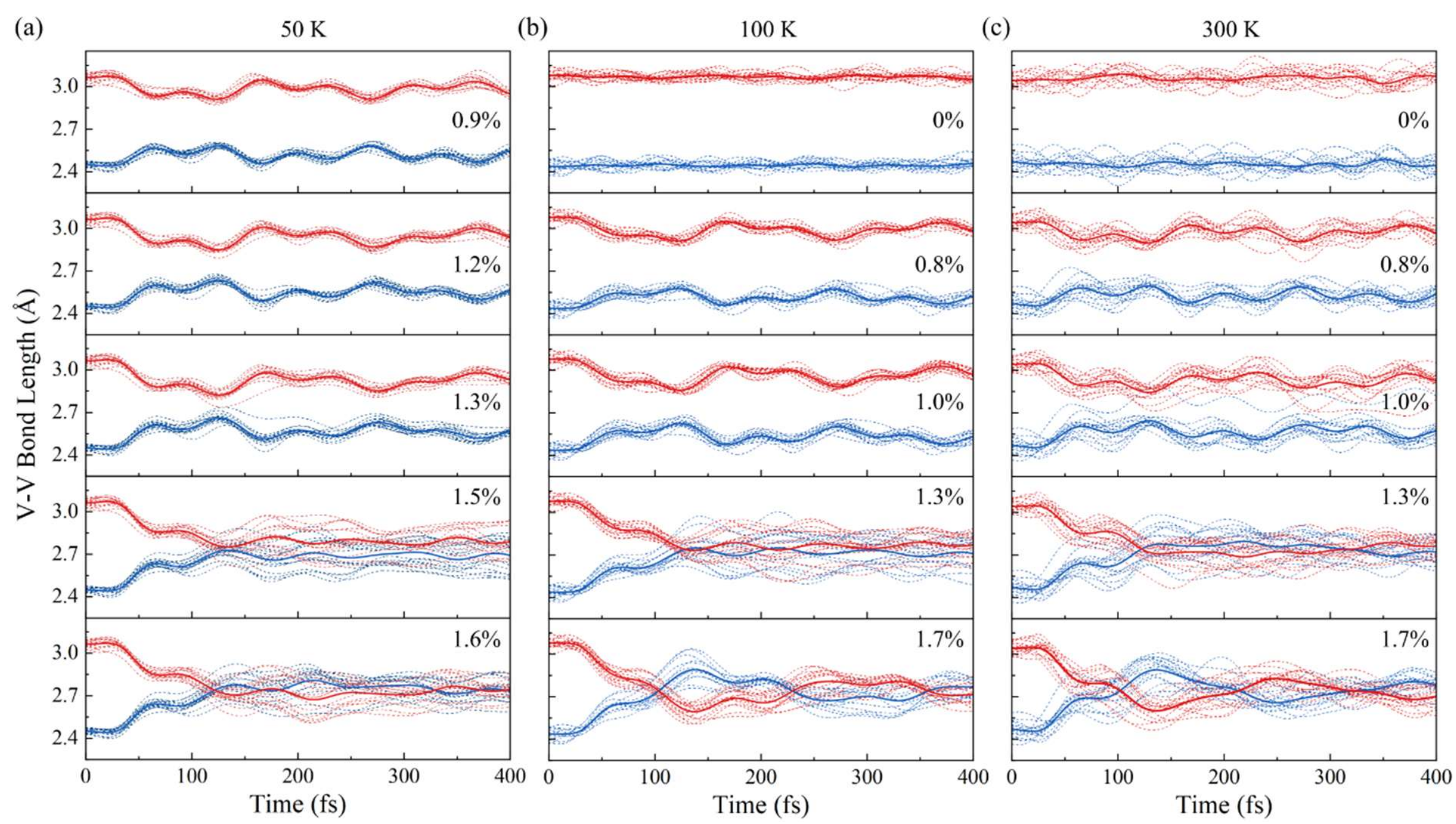

FIG. S3. Temporal dynamics of V-V bond length under different electronic excitations at initial temperature of 50 K (a), 100 K (b), and 300 K (c). The red and blue dashed lines correspond to the longer and shorter bond lengths, respectively. The average bond lengths of longer and shorter bonds are represented by red and blue solid lines.

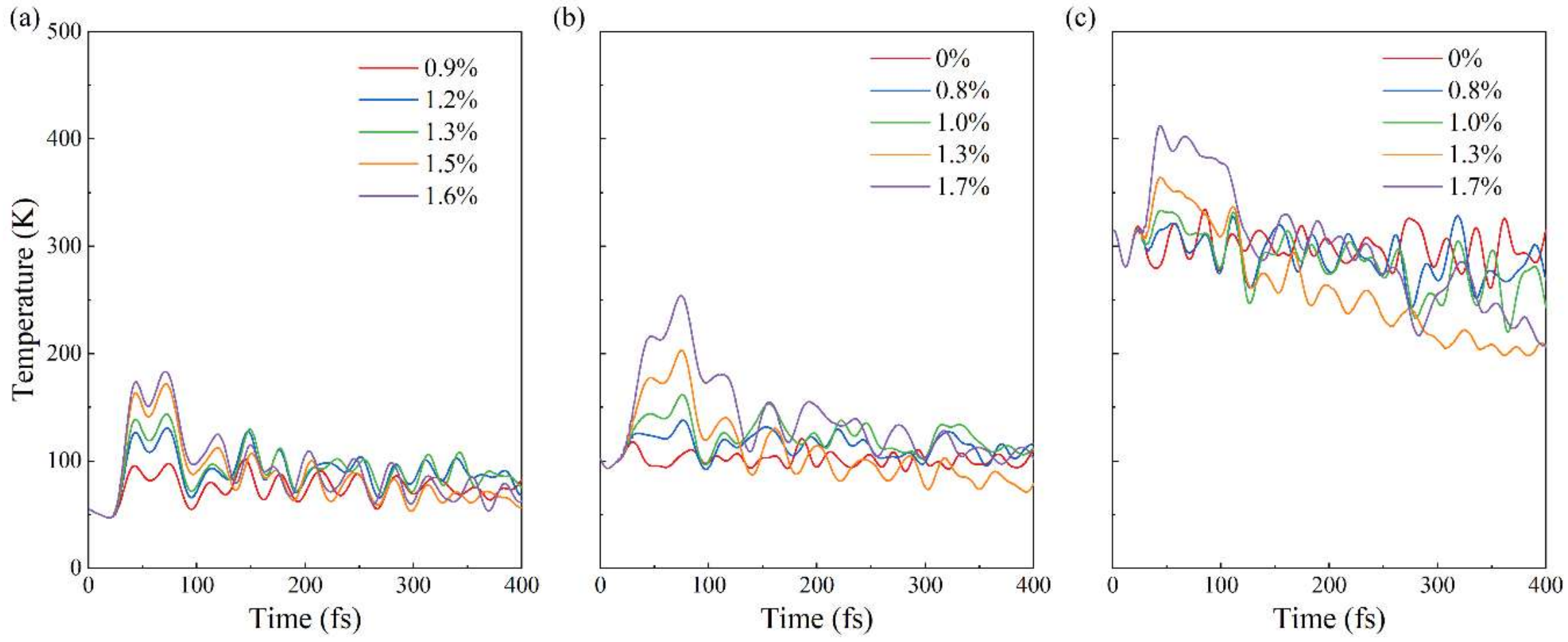

Fig. S4. Evolution of structural temperature during rt-TDDFT simulations under different electronic excitations, with initial temperatures of 50 K, 100 K, and 300 K.

Figure S3 depicts the time-dependent evolution of V-V bond lengths under different electronic excitations at initial temperatures of 50 K, 100 K, and 300 K [Fig. S4]. In the M1 phase, V atoms exhibit two distinct types of V-V bonds, including a set of longer V-V bonds ($d_{\mathrm{I}}$ = 3.21 Å) indicated by the red lines in Fig. S3, and a set of shorter V-V bonds ($d_{\mathrm{II}}$ = 2.51 Å) known as V-V dimers, represented by the blue lines in Fig. S3. In the R phase, all V-V bonds become a same value ($d_{\mathrm{R}}$ = 2.84 Å). For a clear assessment of the pure thermal effects, we also display the evolution of V-V bond lengths without photoexcitation at 100 K and 300 K [Fig. S3]. The longer and shorter bonds both exhibit vibrational fluctuations around their equilibrium positions, with the vibrational amplitude of these bonds being more pronounced at 300 K compared to 100 K. Under a 1.0% electronic excitation, a notable discrepancy emerges in the evolution of bond lengths between 100 K and 300 K. At 100 K, although the short bond lengths exhibit a slight reduction and the long bond lengths undergo a slight increase, the bond lengths remain unequal throughout the dynamics [Fig. S3(b)]. In contrast, at 300 K, a prominent feature is the crossing of partial bond lengths, resulting in the near-equality of short and long bonds at 120 fs [Fig. S3(c)]. At higher excitations (1.3% and 1.5%), the bond lengths converge towards the R phase at 100 K and 50 K, respectively. This analysis reveals that the photoinduced phase transitions in $VO_2$ depends on initial temperature at lower excitations, but the dependence diminishes as excitation increases.

We also calculate the diffraction intensity using the structure factor $F(hkl)$ formula [1,2],

$$F\,(h,k,l) = f_V \sum\nolimits_V \exp[-2\pi i q(hkl)\cdot r_V(t)] + f_O \sum\nolimits_O \exp[-2\pi i q(hkl)\cdot r_O(t)]$$

$$\Delta I(t) = \frac{I\,(t)}{I\,(0)} = \frac{|\langle F(h,k,l)\rangle(t)|^2}{|\langle F(h,k,l)\rangle(0)|^2} \tag{4}$$

where the parameters $f_V$ = 23 and $f_O$ = 8 represent the atomic scattering factor of V and O atoms, respectively. The wave vector is denoted as $q(hkl)$, with a (-122) Bragg peak selected in alignment with experimental measurements [3]. The symbol $r(t)$ signifies the fractional coordinates of either V or O atoms at time $t$ within the unit cell. The structure vectors $F\,(h,k,l)$ are computed at each time step based on the atomic positions obtained from the rt-TDDFT simulations. Given the utilization of a 2×2×2 supercell structure including eight-unit cells, we calculate eight-group $F\ (h, k, l)$ vectors to obtain the final average results. Temporal evolution of the simulated relative intensity $I(t)/I(t\!=\!0)$ from Bragg peak (-122) at 50K, 100 K and 300 K is shown in Fig. 1(b).

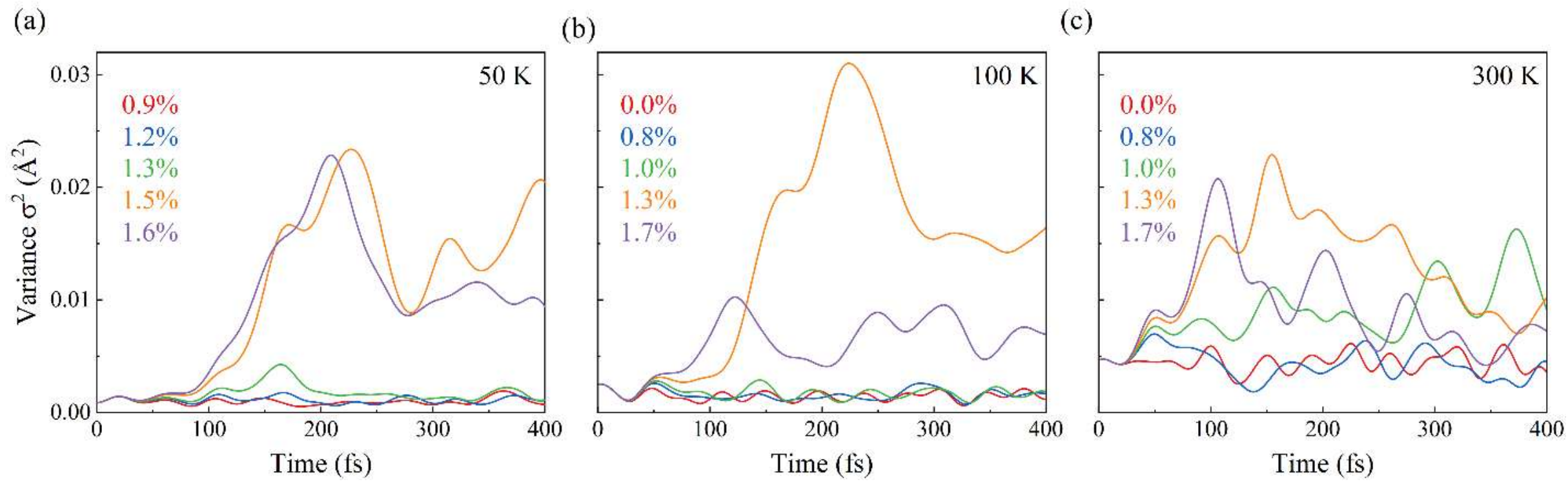

FIG. S5. Time-resolved variance in atomic displacement under different electronic excitations, with initial temperatures of 50 K, 100 K, and 300 K.

For electronic excitations below 0.9%, the variance remains relatively stable, and the variance is 300 K > 100 K > 50 K, indicating a higher degree of atomic disordering at 300 K [Fig. S5]. As the electronic excitation reaches 1.0%, the variance gradually increases at 300 K, signifying the emergence of semiconductor-to-metal transition [4]. However, the variance remains unchanged at 50K and 100 K, indicating the absence of the phase transition. This observation underscores the role of thermal effects stemming from the initial temperature, which effectively reduces the phase transition threshold [5]. The electronic excitation needs to surpass the 1.3% threshold at 100 K and 1.5% threshold at 50 K, and a similar trend of gradual escalation in variance is observed [Figs. S5(a) and S5(b)]. At 1.3% and 1.5% electronic excitations, marking the onset of the phase transition at 50 K and 100 K, the variance undergoes a rapid increase between 50 fs and 200 fs, which results in a larger degree of atomic disordering at 50K and 100 K during the photoinduced dynamics. Similarly, at 1.6% and 1.7% electronic excitations, the degree of atomic disordering at 50 K, 100 K and 300 K experiences a rapid increase between 50 fs and 200 fs, ultimately reaching an equal value of equilibrium after 200 fs. This discovery indicates that atomic disordering triggered by photoexcitation rapidly dissipates memory of the initial thermal distribution near the phase transition point, in concordance with recent experimental observations [3].

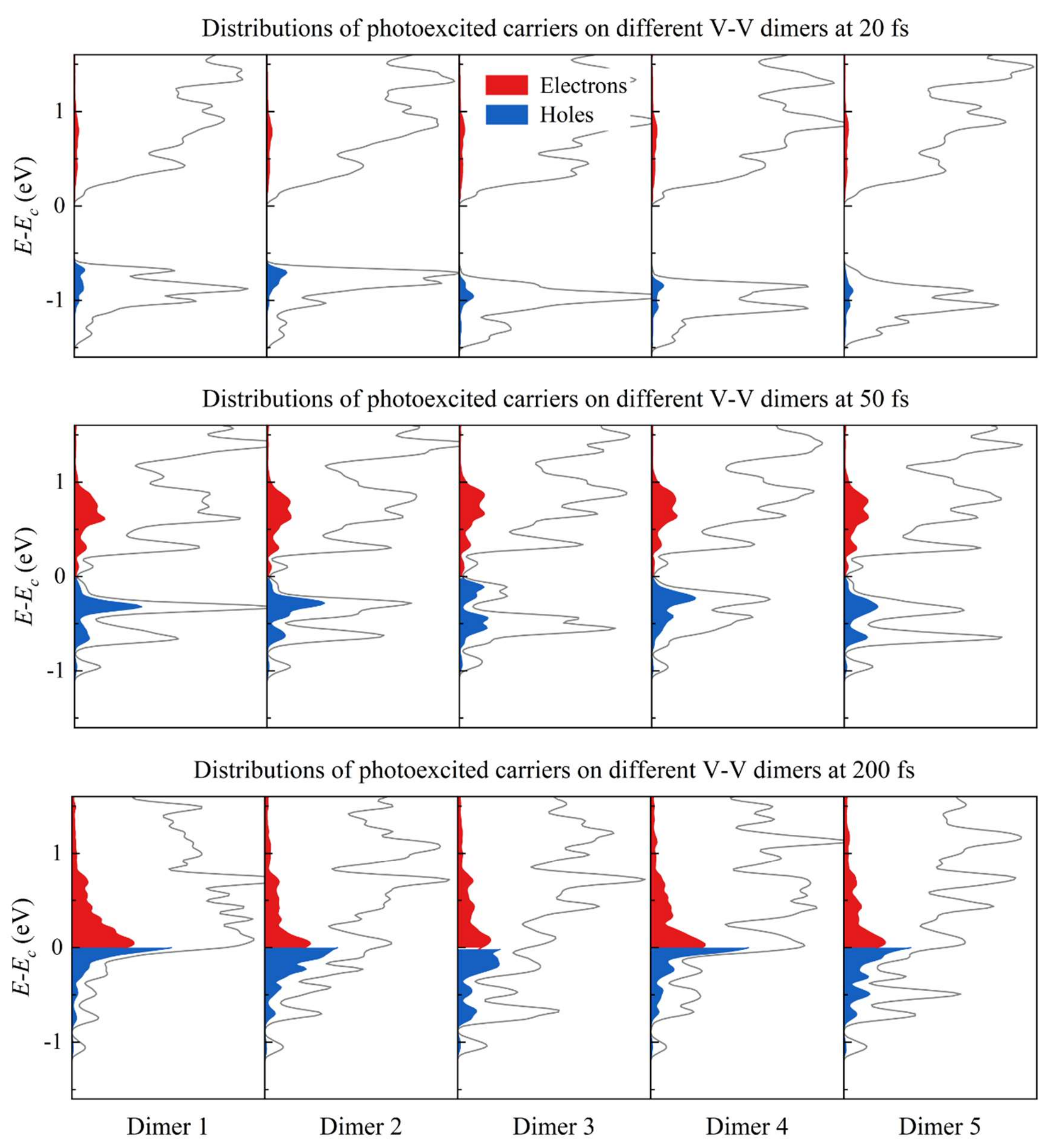

FIG. S6. The projected density of states (PDOS) of different V-V dimers at 20 fs, 50 fs, and 200 fs. The grey lines represent the PDOS of each individual V-V dimer, and the red (blue) shaded regions indicate the distributions of photoexcited electrons (holes).

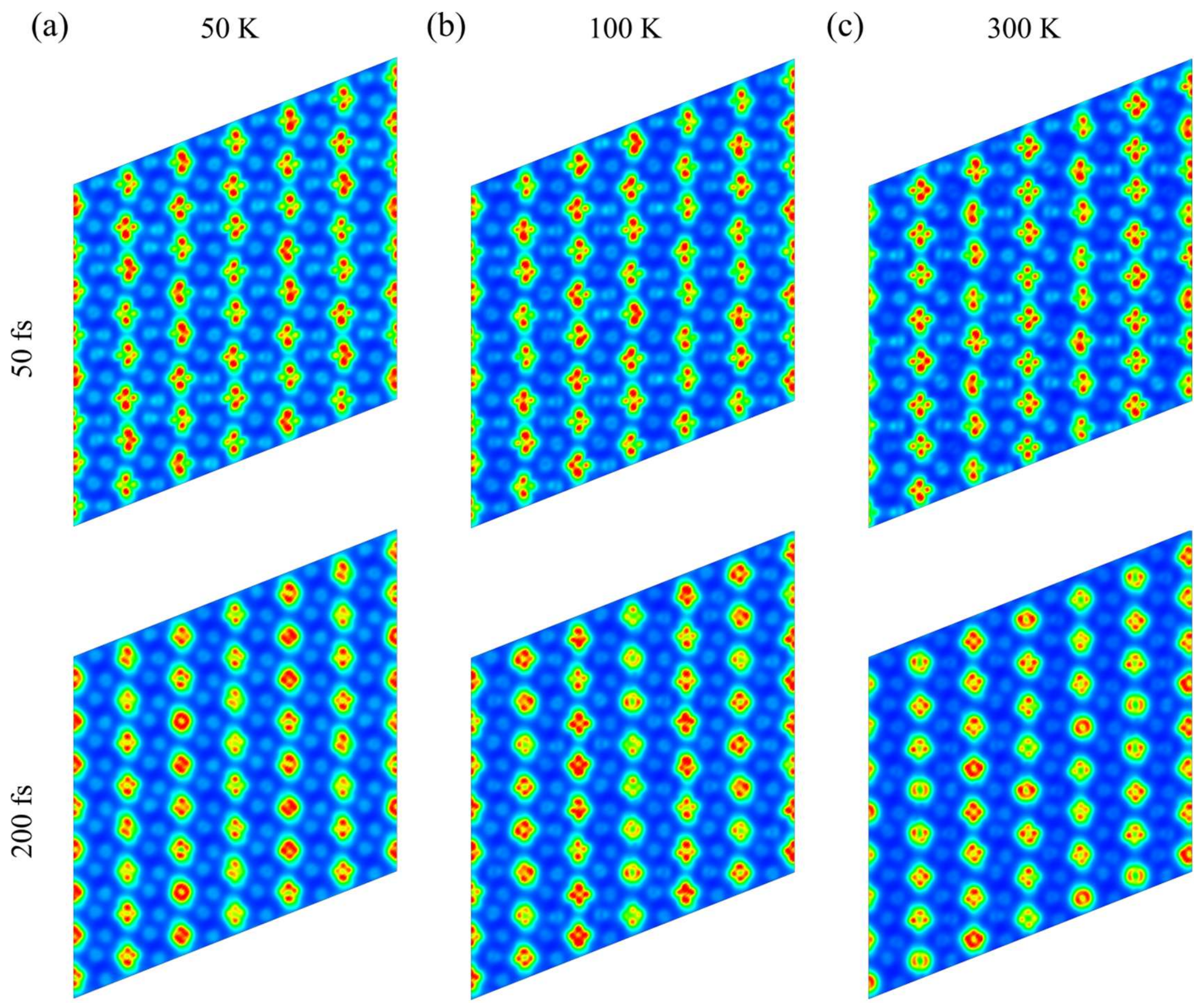

FIG. S7. Real-space distributions of photoexcited holes on the $(0\bar{1}1)$ plane at 50 fs and 200 fs for 50 K (a), 100 K (b), and 300 K (c).

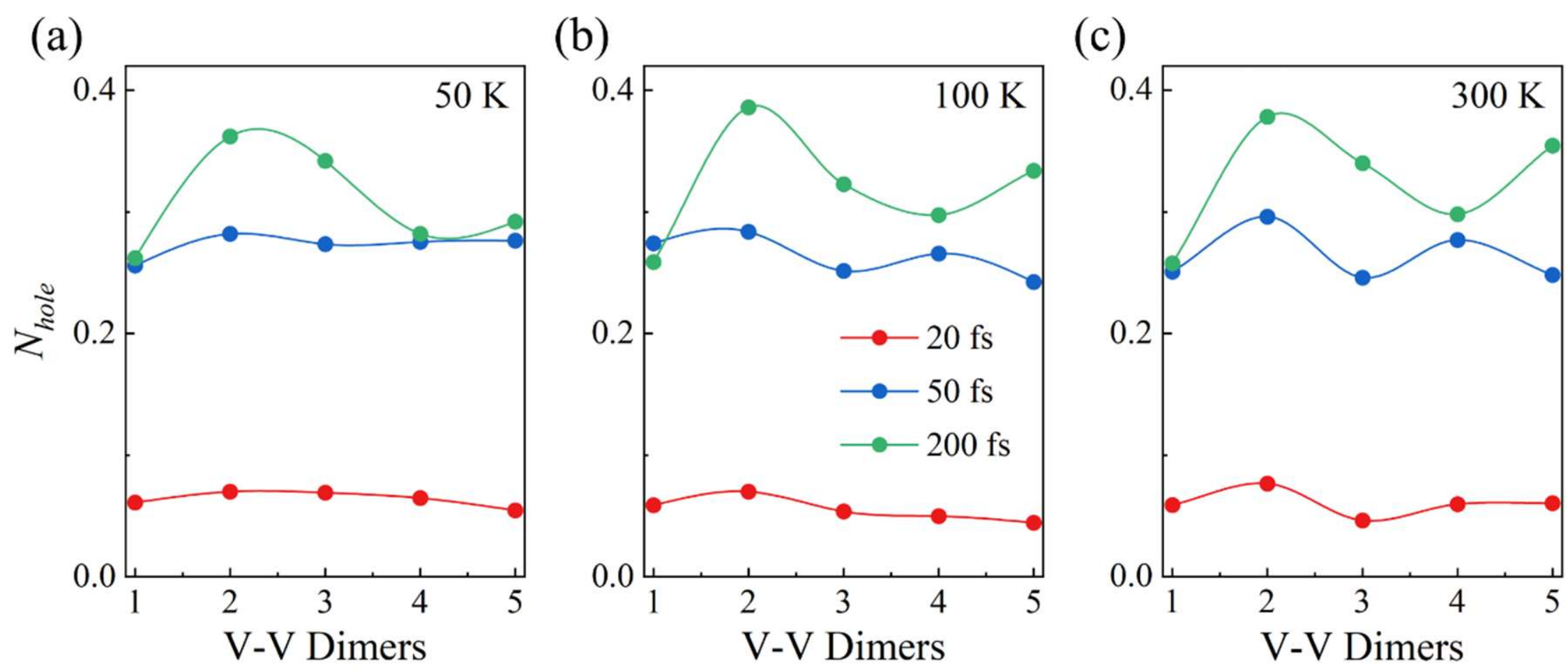

FIG. S8. Photoexcited holes on V-V dimers within -1.6-0 eV below the Fermi Level at 20 fs (red), 50 fs (blue), and 200 fs (green) for 50 K (a), 100 K (b), and 300 K (c).

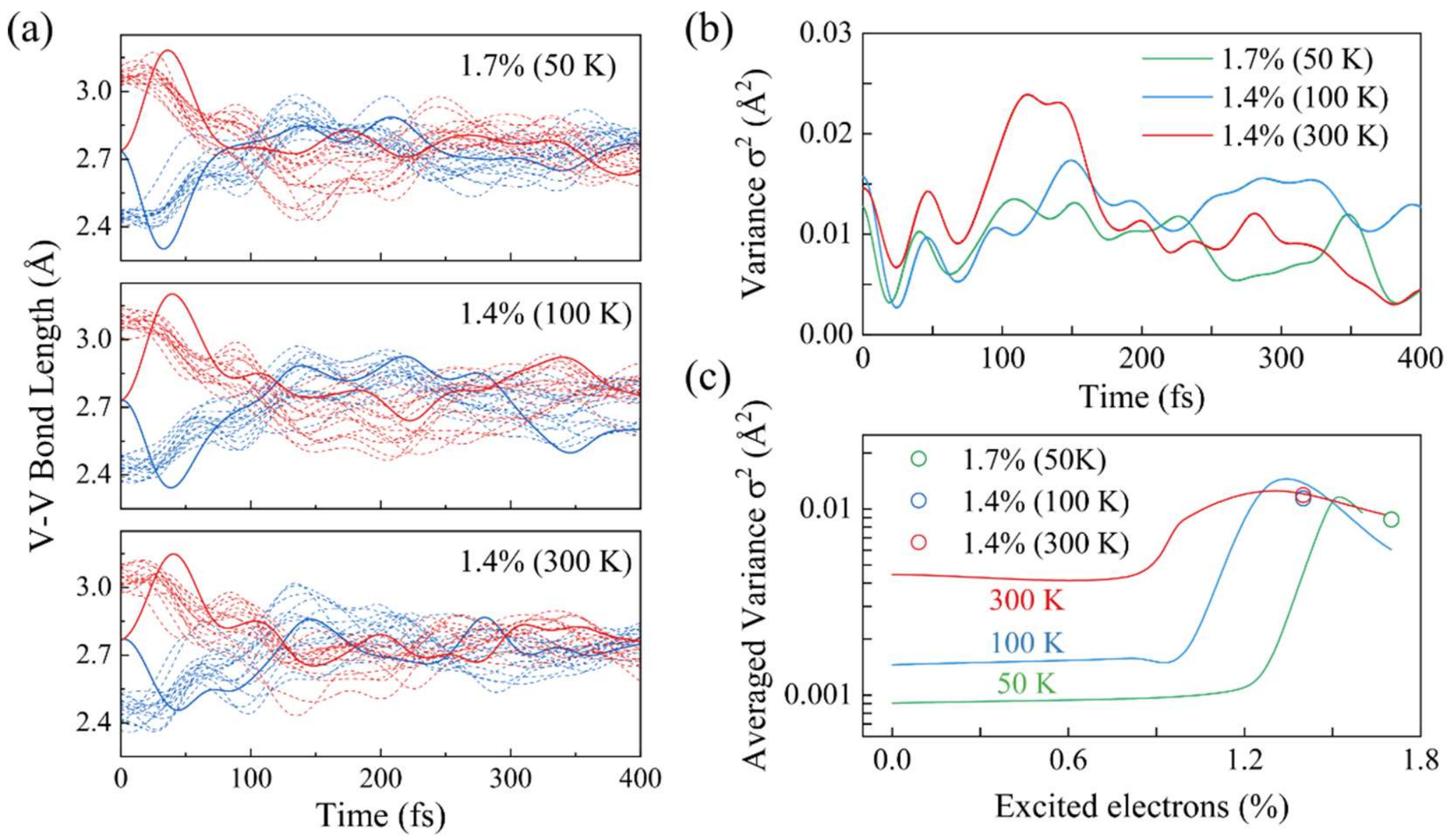

FIG. S9. Influence of an initial polaronic state on atomic disordering. (a) Temporal evolution of the V-V bond length following photoexcitation at 50 K, 100 K, and 300 K. The red and blue dashed lines indicate the longer and shorter bond lengths, respectively. The initially equal V-V bonds are represented by the red and blue solid lines. (b) Time-resolved variance in atomic displacement at 50 K, 100 K, and 300 K. (c) Variance of atomic displacement as a function of the electronic excitation percentage (η) at 50 K, 100 K, and 300 K. The lines are from the results with the initial homogeneous system in Fig. 1(a), and the circles represent the average variance with the initial inhomogeneous state.

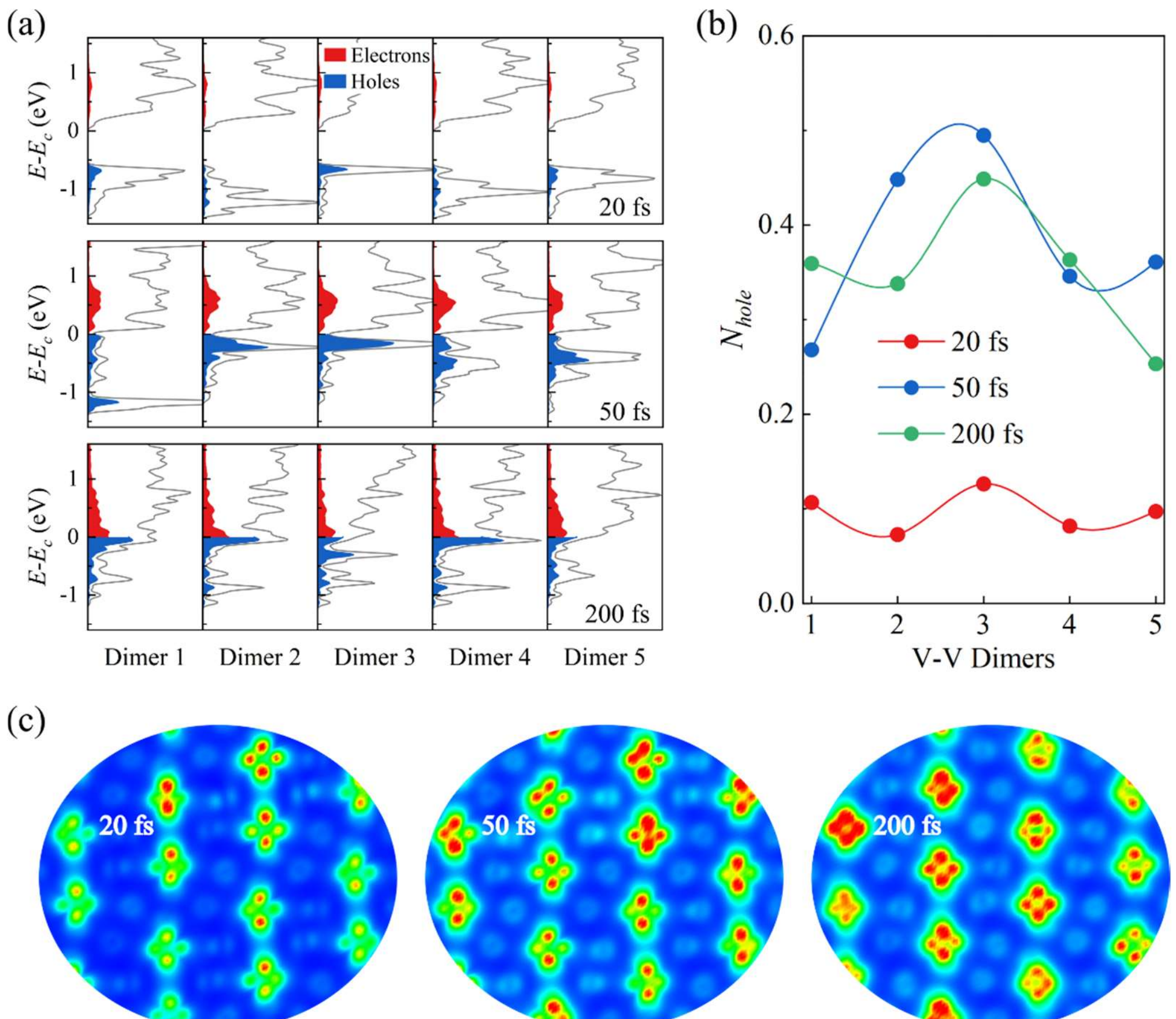

FIG. S10. (a) The projected density of states (PDOS) of different V-V dimers at 20 fs, 50 fs, and 200 fs. The grey lines represent the PDOS of each individual V-V dimer, and the red (blue) shaded regions indicate the distributions of photoexcited electrons (holes). (b) Photoexcited holes on V-V dimers within -1.6-0 eV below the Fermi Level at 20 fs (red), 50 fs (blue), and 200 fs (green). (c) Distributions of the photoexcited holes on the $(0\bar{1}1)$ plane at 20 fs, 50 fs and 200 fs following photoexcitation.

## Note 3. Anisotropy of photoinduced atomic disordering

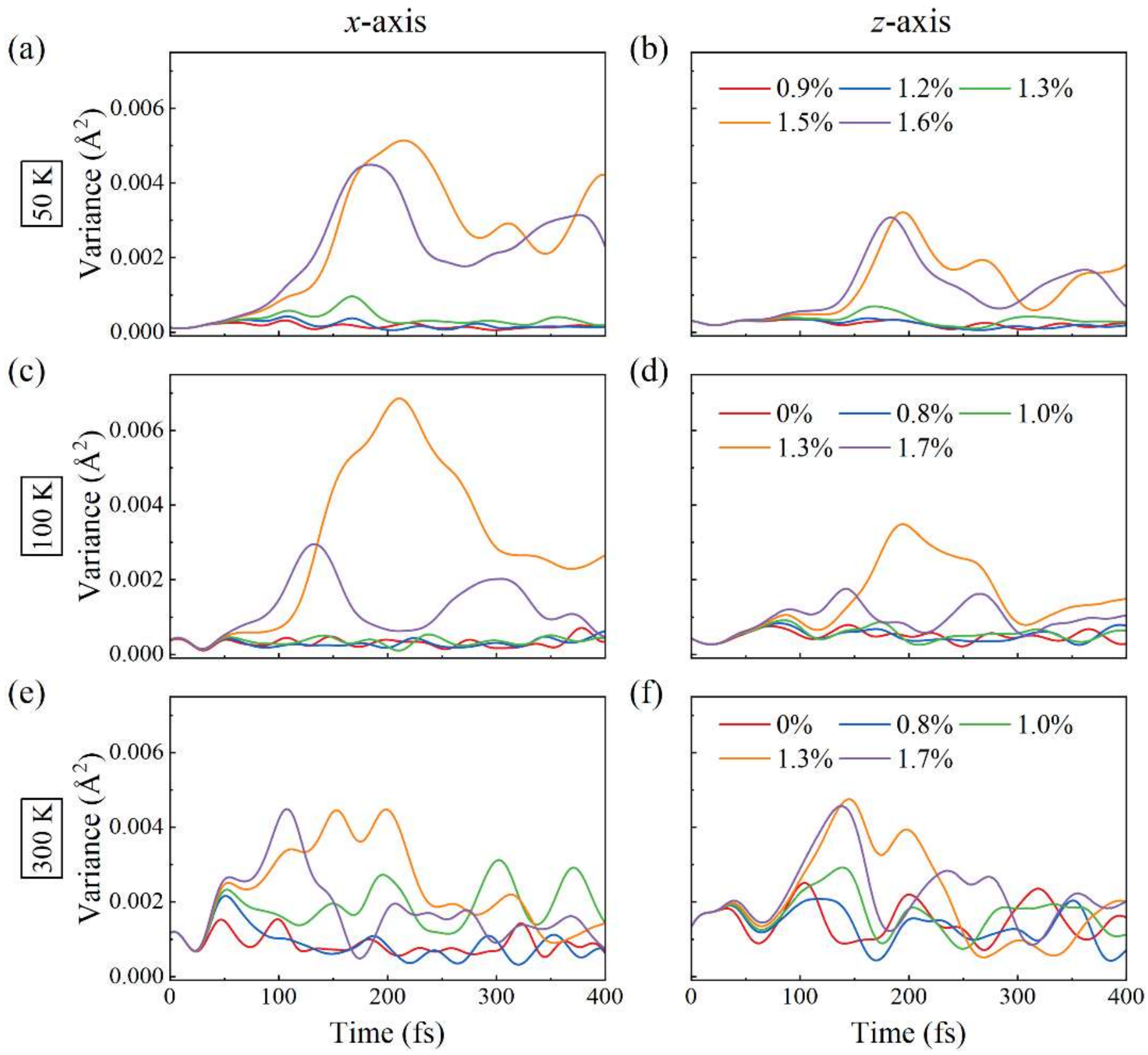

FIG. S11. Time-resolved variance in atomic displacement along *x* and *z* axes under different electronic excitations, with initial temperatures of 50 K, 100 K, and 300 K.

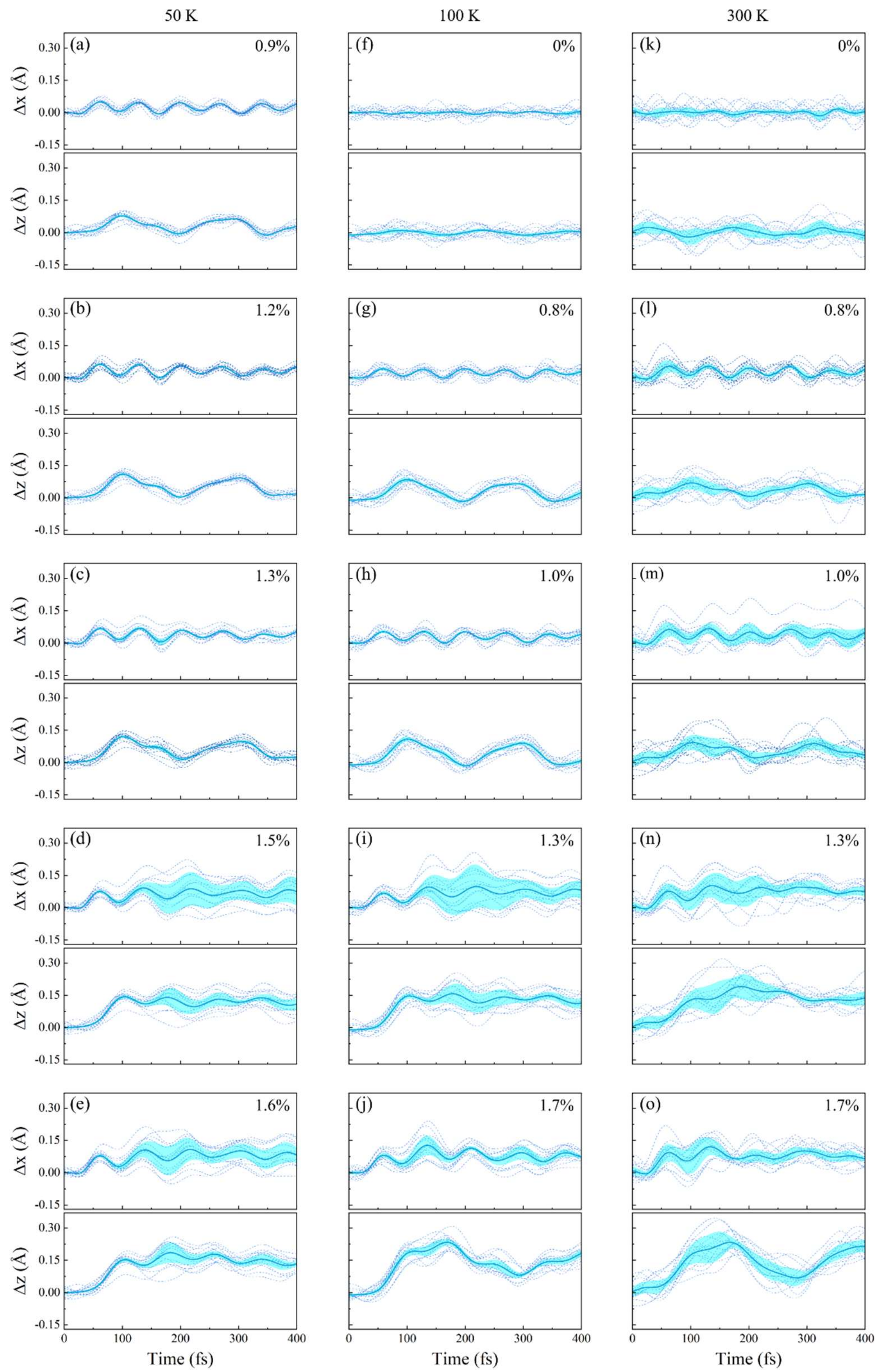

FIG. S12. Temporal evolution of V atomic displacement along *x* and *z* axes for different photoexcitation at initial temperature of 50 K (a-e), 100 K (f-j), and 300 K (k-o). The blue shading indicates the fluctuation of atomic disordering.

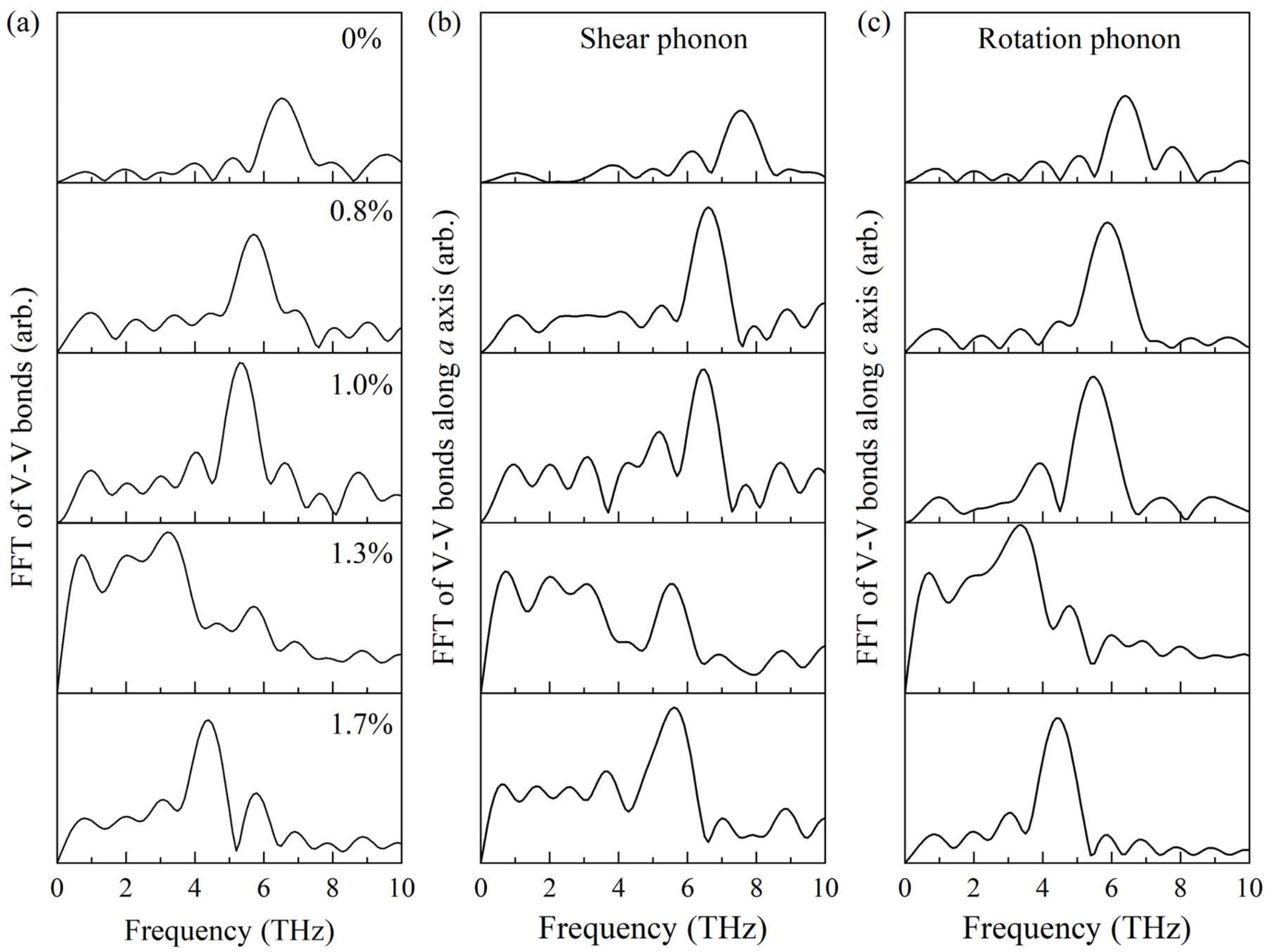

FIG. S13. Phonon modes of V atoms under different photoexcitation. The phonon modes are obtained from the normalized Fast Fourier Transform (FFT) of the average V-V bond lengths: (a) average V-V bond lengths in the overall structure, (b) average V-V bonds bond lengths along the *a* axis, and (c) average V-V bonds bond lengths along the *c* axis.

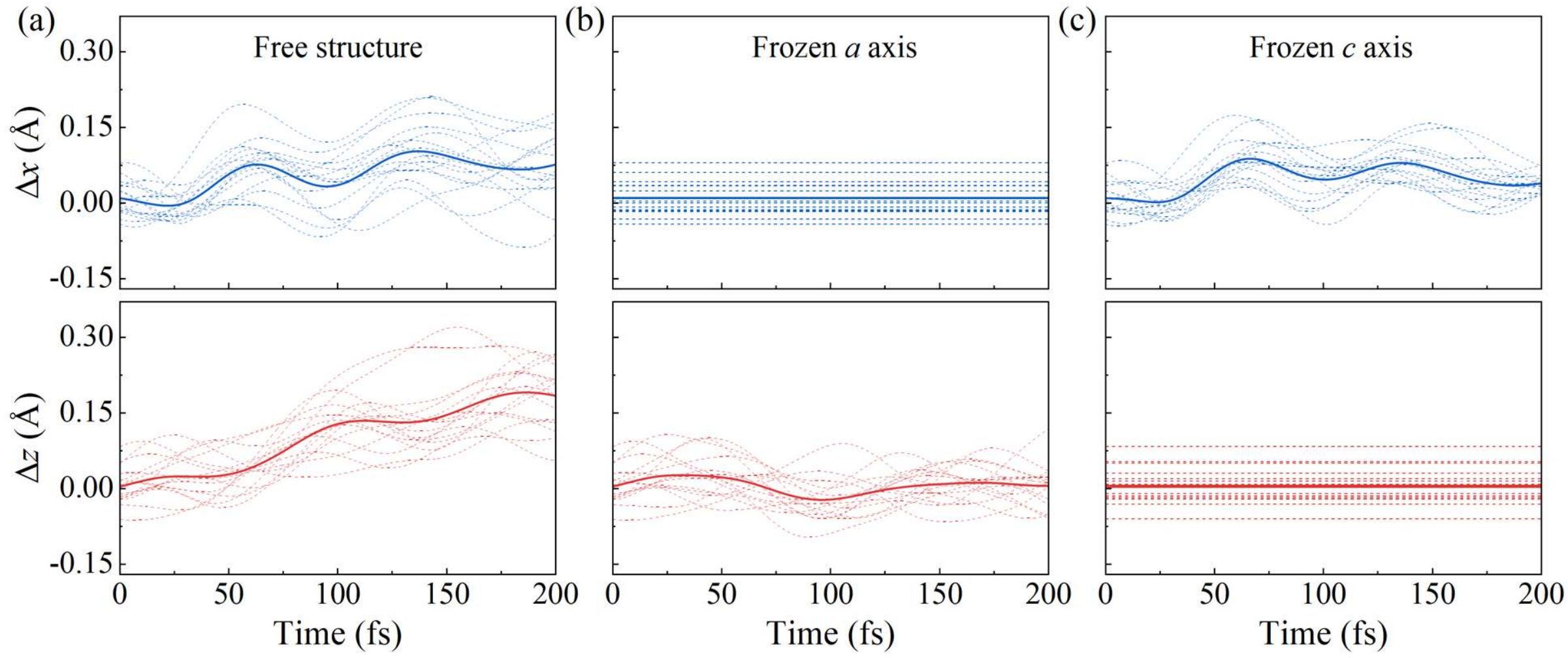

FIG. S14. Temporal evolution of atomic displacement along the *x*- and *z*-axes under 1.3% photoexcitation at 300 K in (a) a free structure, (b) a structure with the *a*-axis frozen, and (c) a structure with the *c*-axis frozen.

Furthermore, we froze the positions of the V atoms along the *x* and *z* directions, respectively. When the positions along the *x* direction are frozen, the V atoms exhibit limited movement along the *z* direction, contrasting with the noticeable changes in the free structure [Fig. S14]. Conversely, when the positions along the *z* direction are frozen, the V atoms maintain substantial motion, similar to that observed in the free structure. This indicates that V atomic motion along the *z* direction is constrained by the displacement along the *x* direction, suggesting that the rotational motion is driven by the initial elongation process.